\documentclass{article}

\PassOptionsToPackage{numbers}{natbib}

 \usepackage[main, final]{neurips_2026}

\usepackage[utf8]{inputenc} %
\usepackage[T1]{fontenc}    %
\usepackage{hyperref}       %
\usepackage{url}            %
\usepackage{booktabs}       %
\usepackage{amsfonts}       %
\usepackage{nicefrac}       %
\usepackage{microtype}      %
\usepackage{xcolor}         %
\usepackage{amsmath}
\usepackage{float}
\usepackage{graphicx}
\usepackage{multirow}
\usepackage[font=small,skip=3pt]{caption}
\usepackage{pifont}

\title{Mechanism-Level Chemical Reaction Simulation with Electron Bookkeeping Transformer}

\author{%
  Sarah Cao$^{1,2}$ \quad
  Shitong Luo$^{3}$ \quad
  Connor W. Coley$^{3,4}$\\[3pt]
  $^1$Department of Chemistry 
  $^2$Department of Mathematics\\
  $^3$Department of Electrical Engineering and Computer Science\\
  $^4$Department of Chemical Engineering\\
  Massachusetts Institute of Technology\\[2pt]
  \texttt{\{syxcao,luost,ccoley\}@mit.edu}
}

\begin{document}

\maketitle

\begin{abstract}
Modeling chemical reaction mechanisms is central to understanding and predicting chemical reactivity. We present Electron Bookkeeping Transformer (\textsc{Bookkeeper}), a data-driven model that simulates chemical reactions at mechanistic resolution by generating electron movement trajectories. Following the symbolic arrow-pushing formalism in chemistry, \textsc{Bookkeeper} explicitly models each electron as a token and predicts its movement sequentially, ensuring mass and charge conservation during both elementary step and reaction pathway predictions by construction. On reaction-size-balanced test sets, \textsc{Bookkeeper} improves top-1 elementary-step accuracy, top-1 pathway rank accuracy, and top-1 to top-5 product prediction accuracy over FlowER. On the entire test set, it achieves higher elementary-step top-1 accuracy but lower pathway accuracy than FlowER. \textsc{Bookkeeper} also samples elementary steps substantially faster, adapts to several unseen reaction types with minimal fine-tuning, and generates candidate chemical transformations for further screening. These results suggest that explicit electron bookkeeping is a useful framework for virtually screening chemical reactions across diverse reagents, validating computer-assisted synthesis planning predictions, and providing mechanistic interpretations of reactivity.

\end{abstract}

\section{Introduction}

Chemists explain transformations from reactants to products in a reaction in terms of their \textit{mechanisms}, which are ordered sequences of elementary steps where electrons move, and bonds form and break. This mechanistic approach is central to reaction development and screening, impurity analysis, and the discovery of new transformations \cite{wen_chemical_2023, margraf_exploring_2023, joung_reproducing_2024, chen_predicting_2025}. Historically, however, machine learning models for reaction prediction are trained to predict only the final major product from a set of reactants, often using SMILES translation, graph edits, or template classification \cite{wei_neural_2016, segler_neural-symbolic_2017, schwaller_molecular_2019, tetko_state---art_2020, jin_predicting_2017, coley_graph-convolutional_2019, dai_retrosynthesis_2019, sacha_molecule_2021, zhong_retrosynthesis_2023}. Although such models can be highly accurate in endpoint prediction, they generally do not explain how products form, often disregard reagents and catalysts, and provide limited information about intermediates, byproducts, or competing pathways compared to those that operate mechanistically \cite{joung_reproducing_2024, tavakoli_ai_nodate, miller_mechanism-aware_2025, das_deepmech_2025}.

Recent efforts to construct mechanistic reaction datasets have made it increasingly feasible to revisit reaction prediction at the mechanism level. In particular, large-scale mechanistic datasets generated from expert-curated elementary reaction templates provide sequences of intermediates between experimentally reported reactants and products, enabling models to predict elementary steps rather than only global outcomes \cite{joung_reproducing_2024, chen_large-scale_2024}. For example, \citet{joung_reproducing_2024} constructed such a dataset by applying expert elementary reaction templates to reactions in the USPTO database, recovering experimentally recorded products and yielding millions of elementary mechanistic steps \cite{joung_reproducing_2024}. Mechanistic models built on similar data have also recovered reaction pathways, identified reagent and catalyst roles, and predicted byproducts or impurities that endpoint models often omit \cite{joung_reproducing_2024, chen_predicting_2025, tavakoli_ai_nodate}. 

However, reaction prediction models can still inherit limitations from their representations. Sequence- and string-based models have flexible input formats, most often operating on SMILES strings, but are often not mass-conserved and can delete or hallucinate atoms \cite{schwaller_molecular_2019, tetko_state---art_2020, neukomm_teaching_2026}. Graph-edit models can conserve atoms through their more constrained data structures, but do not directly model electron flow as the primary generative action and are reported to produce nonphysical steps when bond, hydrogen, and charge edits are predicted independently \cite{sacha_molecule_2021,jorner_machine_2021,jin_predicting_2017,coley_graph-convolutional_2019}. Earlier models such as ELECTRO and NERF explicitly connect reaction prediction to electron movement, but often rely on graph representations of molecules and represent electron flow indirectly as redistributions \cite{bradshaw_generative_2019, bi_non-autoregressive_2021}. In more recent work, Reactron predicts reaction outcomes and accompanying arrow-pushing explanations from reactants, showing the value of using chemical mechanisms as a medium for interpretable generation \cite{chen_predicting_2025}. FlowER further models electron movements as a continuous flow and provides a recent open-source baseline for generative mechanism prediction, but its continuous BE-matrix formulation requires sum-safe rounding after inference to recover chemically valid states \cite{joung_electron_2025}. In contrast, our formulation generates relocations of already-existing individual electron tokens as discrete operations, obviating the need for rounding. The model's representation, sequential state-updating training objective, and inference procedure are distinct from FlowER, while both build on previous electron flow modeling.

We thus introduce Electron Bookkeeping Transformer (\textsc{Bookkeeper}), a generative transformer for chemical reaction mechanism prediction. 
The model represents a mechanistic state as a set of atoms and a multiset of valence electrons, where each electron token is assigned to either a single atom as a lone electron or between two atoms as a bonding electron.
An elementary arrow-pushing action moves one existing electron from a source location to a destination location, and repeating such actions updates the electron bag and produces a sequence of electron movements.
Because generation moves existing electron tokens over a fixed atom set, \textsc{Bookkeeper} preserves both atoms and the total number of electrons by construction.
Empirically, \textsc{Bookkeeper} improves elementary-step prediction and top-1 through top-5 pathway recovery on reaction-size-balanced test subsets relative to FlowER \cite{joung_electron_2025} and MechSMILES T5 \cite{neukomm_teaching_2026}. On the full test set, however, FlowER achieves overall higher pathway accuracy, while \textsc{Bookkeeper} retains a smaller top-1 elementary-step advantage. \textsc{Bookkeeper} also offers $>100\times$ faster elementary-step sampling, which is useful for large-scale reaction screening or exploration of deep or broad mechanistic pathways. 
We also find that \textsc{Bookkeeper} performs similarly to FlowER on out-of-distribution reactions and can be effectively fine-tuned with minimal additional data.
Furthermore, we conduct combinatorial virtual reaction screening following \citet{chen_predicting_2025} to demonstrate that our approach can predict novel reactivity as well as extrapolate known mechanisms.
We also examine predictions of representative regioselectivity examples and responses to stoichiometric perturbations, suggesting that our model can learn chemically meaningful preferences.

Our contributions can be summarized as follows:
\begin{itemize}
    \item We introduce Electron Bookkeeping Transformer (\textsc{Bookkeeper}), which represents electrons as tokens and generates chemical reaction mechanisms as sequences of electron movements.  This tokenized representation preserves mass and charge conservation by construction and produces interpretable arrow-pushing trajectories.
    \item We characterize the accuracy of \textsc{Bookkeeper} on a large mechanistic dataset used in prior work and reaction-size-balanced subsets, identifying strengths in top-ranked elementary-step prediction as well as limitations in full-pathway recovery. Its generative architecture enables efficient elementary-step sampling and complete trajectory generation without beam search.
    \item We demonstrate \textsc{bookkeeper}'s adaptation to out-of-distribution reaction types and corroborate its chemically meaningful behavior through combinatorial reaction discovery, regioselectivity, and robustness to excess reagents.
\end{itemize}

\section{Related work}
\paragraph{Mechanisms in chemical reaction prediction} Machine-learning methods for reaction prediction have typically focused on predicting the final major product from a set of reactants and reagents. Approaches in the past decade have posed reaction prediction as template selection, reaction-center identification, or learned chemical reactivity prediction \cite{wei_neural_2016, coley2017prediction, segler_neural-symbolic_2017}. Later models improved flexibility by representing reactions as SMILES-to-SMILES translation problems, graph-to-sequence problems, molecular graph edits, or related graph-based formulations for retrosynthesis and synthesis planning \cite{jin_predicting_2017, coley_graph-convolutional_2019, schwaller_molecular_2019, tetko_state---art_2020, sacha_molecule_2021}. These endpoint-prediction models have converged to high accuracies on typical benchmark datasets, but they abstract away the elementary steps by which transformations occur, providing little information about intermediates, byproducts, or the mechanistic basis of selectivity. 

Mechanistic reaction prediction models address this limitation by predicting sequences of elementary intermediate states rather than only final products \cite{joung_reproducing_2024, chen_predicting_2025}. \textsc{No Electron Left Behind} first demonstrated the value of mechanistic, electron-based rules before the current deep-learning era \cite{chen_no_2009}. More recent models and representations such as RMechRP, DeepMech, and MechSMILES predict radical mechanisms through contrastive learning over reactive atom or orbital pairs \cite{tavakoli_ai_nodate}, use mechanistic supervision for interpretable stepwise prediction \cite{das_deepmech_2025, miller_mechanism-aware_2025}, or formulate arrow-pushing mechanisms into text to fine-tune large language models \cite{neukomm_teaching_2026}. Since balanced chemical reactions can be fully described by changes in electron distribution among atoms of reacting molecules, some mechanistic approaches further focus on modeling electron movement: ELECTRO introduced a generative model of pseudo-electron paths \cite{bradshaw_generative_2019}, predictors of electron redistribution and movement represent reactions through changes in electron or bond structure \cite{bi_non-autoregressive_2021, chen_predicting_2025}, and FlowER formulates mechanism prediction as electron flow matching over continuous bond-electron matrices \cite{joung_electron_2025}. Notably, Reactron \cite{chen_predicting_2025} predicts reaction mechanisms by iteratively generating electron-pair movements between source and sink sites, updating the molecular state after each predicted electron path. \textsc{Bookkeeper} similarly generates reactions through successive electron-movement predictions. Unlike Reactron, \textsc{Bookkeeper} explicitly represents individual valence electrons as tokens, updates their locations one movement at a time, and employs shuffled-prefix training to accommodate the unordered nature of electron movements within an elementary step. 

\paragraph{Autoregressive and discrete generative modeling} Autoregressive generative models have been widely used in language modeling \cite{bengio_neural_nodate, sutskever_sequence_2014, vaswani_attention_2017, you_graphrnn_2018}, and related trajectory-modeling approaches similarly treat decision processes as sequences of actions \cite{chen_decision_2021, janner_offline_2021}. Discrete generative modeling has also been studied through permutation-aware training, denoising, masked generation, and constrained decoding \cite{yang_xlnet_2019, austin_structured_2021, he_diffusionbert_2023, sahoo_simple_nodate, koo_automata-based_2024}. This is especially chemically relevant, as reaction mechanisms do not have a unique canonical arrow order when multiple electrons move in a concerted manner (i.e., simultaneously) within a single mechanistic step. To account for this, \textsc{Bookkeeper} is trained on shuffled partially completed electron-flow states, encouraging mechanism completion without relying on one arbitrary total ordering of arrows. During inference, we use frequency-ranked stochastic sampling instead of relying on raw model scores \cite{holtzman_curious_2020, mihaylova_scheduled_2019}.

\section{Methods} \label{Methods}

\subsection{Formulation}
\textbf{Representation for molecules} A \emph{structure} in a mechanistic reaction state includes the set of atoms present in a reaction \(A = \{a_1,\ldots,a_n\}\), where each atom is defined by its element type and indexed. The set of atoms is conserved through any prediction, as mechanisms may move electrons and change bonding but do not create or delete atoms. A valence electron in a reaction is either part of a lone pair or radical on a single atom or a bonding electron shared between two atoms, so we represent each electron by its location \(\ell = (i,j)\)  with \(i,j \in \{1,\ldots,n\}\), where \(i=j\) denotes an electron localized on atom \(a_i\), and \(i\neq j\) denotes an electron shared between atoms \(a_i\) and \(a_j\). Since electrons in covalent bonds are shared symmetrically, \((i,j)\) and \((j,i)\) represent electrons in the same set of bonds. During training, \(i,j\) for each electron are randomly swapped with probability \(0.5\), augmenting both equivalent endpoint orderings. An \emph{electron bag} \(E \in \mathcal{M}(\mathcal{L})\) is thus a multiset denoting a mechanistic electron configuration over electron locations, where \(\mathcal{L}\) is the space of possible electron locations over the atom set. The multiplicity of a location records how many electrons occupy that lone or bonding position. For example, two electrons between atoms correspond to a single bond, four to a double bond, and six to a triple bond.

\textbf{Updates to reaction states} Any state \(S\) can therefore be represented as \(S = (A,E)\), denoting a specific discrete distribution of electrons in \(E\) over \(A\). An \emph{arrow} \(u = (\ell_{\mathrm{src}}, \ell_{\mathrm{dst}})\) denotes an arrow-pushing action that moves one electron from a source location \(\ell_{\mathrm{src}}\) to a destination location \(\ell_{\mathrm{dst}}\). Each arrow updates the electron bag by removing one electron from the source and adding one electron to the destination, yielding \(E' = E - \{\ell_{\mathrm{src}}\} + \{\ell_{\mathrm{dst}}\}\). The atom set is unchanged, so the updated state is \(S' = (A,E')\).

\textbf{Mass-conserved reaction steps and trajectory} An \emph{elementary step} is an unordered set of arrows \(\tau = \{u_1,\ldots,u_K\}\) that results in the next reaction state \(S'\), where \(K\) is typically between 1 and 6. A mechanistic \emph{reaction pathway} \(\Pi\) is a finite sequence of elementary steps, represented as a trajectory of states \(S_0 \xrightarrow{\tau_1} S_1 \xrightarrow{\tau_2} \cdots \xrightarrow{\tau_T} S_T.\) Because each update removes and adds exactly one electron and the atom set is conserved, \(|E_t| = |E_0|, A_t = A_0\) for all \(t\). The numbers of atoms and electrons are invariant by construction.

\textbf{Chemical validity} We define a generated elementary step to be chemically valid if all its arrows can be applied to the current electron bag and the resulting electron configuration satisfies valence and bond formation rules for each atom. A generated pathway is chemically valid if every generated elementary step yields a valid molecular state.

\subsection{Autoregressive transformer model to predict electron movements}
\label{Methods:model}
\begin{figure}[t]
  \centering
  \makebox[\textwidth][c]{%
    \includegraphics[width=1\linewidth]{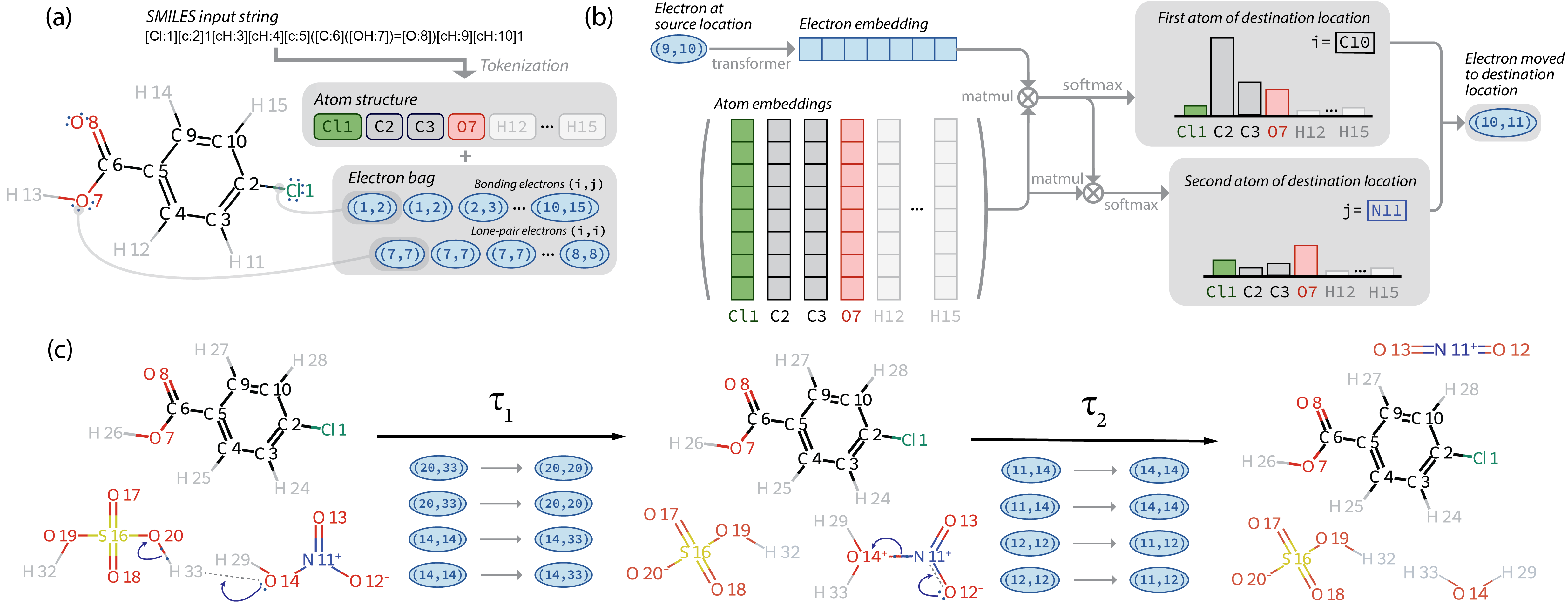}%
}
  \caption{
Electron bookkeeping formulation for mechanism-level reaction simulation. 
(a) \textsc{Bookkeeper} converts an atom-mapped molecular state in SMILES format into an atom structure and an electron bag.
(b) Given the current electron bag, \textsc{Bookkeeper} predicts an electron movement in stages through a transformer architecture by selecting a source electron and then predicting the first and second atom indices of the destination location.
(c) Example mechanistic steps in a reaction pathway for an electrophilic aromatic substitution reaction.}
  \label{fig_1}
\end{figure}
\textsc{Bookkeeper} defines a stochastic policy over electron-movement actions. Given a current mechanistic state \(S_t=(A,E_t)\), the model samples one electron movement at a time, applies the movement to the electron bag, and repeats until an elementary step is completed. Thus, an elementary mechanism with \(K\) arrows is generated as \(p_\theta(\tau \mid S_0)=\prod_{k=1}^{K}p_\theta(u_k \mid S_{k-1})\), with \(S_k = \operatorname{Apply}(S_{k-1},u_k)\), \(S_{0}=S_t\), and \(S_{K}=S_{t+1}\). Atom and electron representations are tokenized, and each electron is additionally appended with a copy index that distinguishes multiple electrons occupying the same location. Tokens are then concatenated and passed through a transformer encoder \cite{vaswani_attention_2017}, allowing electron tokens to attend to atoms, other electrons, and the current global mechanistic context.

The model predicts an electron movement in three sequential stages. First, it predicts a \emph{gate} distribution over all current electron tokens plus a stop action, \(p_\theta(g_k \mid S_{k-1})\) with \(g_k \in \{1,\ldots,|E_{k-1}|,\mathrm{stop}\}\). The gate determines whether generation should continue and, if so, which source electron should move next. If the stop action is selected, no additional electron movement is generated for the current elementary step. If an electron token is selected, that token defines the source electron \(e^{\mathrm{src}}_k\) and its current source location \(\ell^{\mathrm{src}}_k\). Second, conditioned on the selected source electron and the current state, the model predicts the first atom index of the destination location with \(p_\theta(i_k \mid e^{\mathrm{src}}_k,S_{k-1})\). Third, conditioned additionally on this first destination atom, the model predicts the second atom index with \(p_\theta(j_k \mid e^{\mathrm{src}}_k,S_{k-1},i_k)\). 

Equivalently, the model factorizes the electron-movement probability as
\[
p_\theta(u_k \mid S_{k-1})
=
p_\theta(e^{\mathrm{src}}_k \mid S_{k-1})
p_\theta(i_k \mid e^{\mathrm{src}}_k,S_{k-1})
p_\theta(j_k \mid e^{\mathrm{src}}_k,S_{k-1},i_k)
\]
The electron bag is then updated with the sampled action to arrive at the next state \(S_k=(A,E_k)\), from which the model samples the next action.

\subsection{Training}
Electron movements are considered simultaneous in the same elementary step. We train \textsc{Bookkeeper} with shuffled-prefix supervision to avoid introducing unnecessary order dependence among arrows in the same step. During training, for an elementary step with ground-truth arrows \(\tau = \{u_1,\ldots,u_K\}\), we sample a partial mechanistic step ("prefix") and require the model to generate a valid continuation. Specifically, we choose a random ordering \(\pi\) of the arrows, sample a prefix length \(m\), apply the prefix actions to the initial electron bag, and train the model from the resulting partial state \(S_m = \operatorname{Apply}(S_0, (u_{\pi(1)}, \ldots, u_{\pi(m)}))\). The remaining actions \(R_m = \{u_{\pi(m+1)},\ldots,u_{\pi(K)}\}\) define the set of valid next generative decisions from \(S_m\). 

The model is then trained using the three-stage action factorization described in the previous section: gate/source selection, first destination atom prediction, and second destination atom prediction. During training, the gate distribution is supervised with a multi-positive target over all remaining ground-truth source electrons. Let \(
\mathcal{G}_{\mathrm{rem}}(S_m)=\{g(\ell_{\mathrm{src}}):(\ell_{\mathrm{src}},\ell_{\mathrm{dst}})\in R_m\}\) denote the set of electron-token indices corresponding to remaining source locations in the current electron bag. If \(R_m\neq\emptyset\), every element of \(\mathcal{G}_{\mathrm{rem}}(S_m)\) is treated as a valid positive gate choice. If \(R_m=\emptyset\), the stop class is the only positive gate target. The gate loss is therefore
\(\mathcal{L}_{\mathrm{gate}}=-\sum_{g \in \mathcal{G}_{\mathrm{rem}}(S_m)}\log p_\theta(g \mid S_m),\) where \(g=\mathrm{stop}\) when no arrow remains. This multi-positive gate objective allows any remaining valid electron movement to be selected next, rather than forcing the model to predict one arbitrary next arrow.

For each moved electron, the destination objective supervises the two atom indices defining its target electron location. If a remaining arrow has source gate index \(g \neq \mathrm{stop}\) and destination location \((i,j)\), the destination contribution gives objective \[\mathcal{L}_{\mathrm{dst}}=-\sum_{(g,(i,j))\in R_m}\left[\log p_\theta(i \mid g,S_m)+\log p_\theta(j \mid g,i,S_m)\right]\] 

Therefore, the full training objective is \(\mathcal{L} =\mathcal{L}_{\mathrm{gate}}+\mathcal{L}_{\mathrm{dst}}\). By exposing the model to many shuffled partial mechanistic steps, shuffled-prefix training encourages generation without reliance on a single imposed arrow order.

\subsection{Generative prediction of elementary steps and reaction pathways}
\paragraph{Elementary step sampling} At inference time, \textsc{Bookkeeper} generates mechanistic elementary steps by stochastic autoregressive sampling. Given an initial state \(S_0\), the model repeatedly samples a source electron and destination location and updates the electron bag with the predicted arrow. Because the arrows within an elementary step are modeled sequentially even though they collectively describe a single chemical transformation, intermediate electron configurations may not correspond to complete molecular states. We therefore identify completed elementary steps using chemical integrity criteria. Specifically, after each sampled arrow update, we examine atoms touched by the generated arrows and require that: (i) each touched atom has nonzero associated electrons, (ii) each touched atom has an even number of bonding electrons, so that bonds correspond to paired-electron bond orders, and (iii) the total number of lone plus bonding electrons around each touched atom does not exceed its valence-shell capacity. Since the model also predicts a stop action as described in Section \ref{Methods:model}, elementary-step prediction terminates either when the stop action is predicted or when it passes these integrity checks, whichever happens first. This procedure ensures that extracted elementary steps satisfy the implemented chemical constraints, without guaranteeing physical or kinetic feasibility. Generated samples that contain trivial reverse loops, where an electron movement is later immediately reversed by the opposite arrow, are discarded.

\paragraph{Pathway generation} We can directly generate complete reaction pathways by repeatedly sampling arrows as described above until the model predicts the stop action for all electrons or the maximum iteration limit terminates the trajectory. Using this method, the model samples an electron trajectory \(\widehat{\Pi}\) from overall reactants in state \(S_0\) to overall predicted products in \(\widehat{S}_T\), through a series of arrows describing a sequence of elementary steps. To extract intermediate states \(\{S_1,...,S_{T-1}\}\) from the trajectory, we apply the same chemical validity criteria checks as above to identify elementary steps.

Besides the natural sequential trajectory formulation that predicts an entire mechanism end-to-end as a consecutive sequence of arrows, \textsc{Bookkeeper} is also compatible with other inference algorithms such as beam search. At each pathway depth, the model generates and frequency-ranks candidate next elementary steps, and beam search retains a set of partial pathways for further expansion. Thus, the same elementary-step generation process can be broadly used as a local generative transition model for broader pathway search.

\section{Experiments} \label{Experiments}
We design experiments to address four aspects of \textsc{Bookkeeper}: First, following \citet{joung_electron_2025}, we evaluate its in-distribution performance on both top-\(k\) elementary-step prediction and top-\(k\) reaction pathway prediction, as well as demonstrate its higher inference speed compared to benchmark models. Second, also following \citet{joung_electron_2025}, we evaluate its out-of-distribution generalization and effectiveness of data-efficient fine-tuning. Third, following \citet{chen_predicting_2025}, we examine its application to combinatorial reaction discovery to assess whether the model can predict plausible and diverse reactivity from enumerated reactant pools. Fourth, we probe whether \textsc{Bookkeeper} learns chemically meaningful principles by investigating its regioselective behavior and robustness to excess reagents.

\subsection{Performance} \label{Performance}
We evaluate \textsc{Bookkeeper} on two in-distribution test settings: reaction-size-balanced subsets and the entire FlowER test set. The full FlowER test set contains 28,049 unique chemical reactions. To facilitate evaluation with end-to-end trajectory sampling, we construct smaller subsets of 1,010 reactions balanced across six atom-count bins, increasing the representation of larger reactions relative to the full test distribution. We compare against FlowER \cite{joung_electron_2025} and MechSMILES T5 \cite{neukomm_teaching_2026}, the two closest open source baselines for generative mechanistic prediction. FlowER is a continuous BE-matrix electron-flow model, while MechSMILES T5 represents mechanisms as textual arrow-pushing sequences, then uses this formulation to train a T5 encoder-decoder model \cite{rael_exploring_nodate}. The balanced-subset comparisons use the released baseline checkpoints of both benchmarking models on the mechanistic dataset of Joung et al. \cite{joung_electron_2025}; the full test set results compare \textsc{Bookkeeper} with FlowER.

\paragraph{Elementary-step prediction}
For elementary-step prediction, each model is given a mechanistic state \(S_t\) and asked to predict the next state \(S_{t+1}\). For \textsc{Bookkeeper}, we sample 64 candidate elementary steps, convert each resulting electron bag into a molecule, canonicalize the predicted molecule, group duplicate predictions, and rank candidates by empirical frequency. A prediction is counted as correct if the canonicalized SMILES of the next intermediate or product after one generated elementary step exactly matches the ground-truth next intermediate or product. 

\paragraph{Pathway accuracy}
Full-pathway accuracy evaluates the prediction bottleneck of the ground-truth elementary steps along a reference pathway \cite{joung_electron_2025}. Following the pathway accuracy definition of FlowER, for a reaction with $N$ different reference pathways, let $L_i$ be the length of the $i$-th pathway, and $r_{ij}$ be the rank of the ground truth predicted from the starting state at step $j$ of pathway $i$. The rank of the reaction is $R = \min_{i=1,\ldots,N}\max_{j=1,\ldots,L} r_{ij}$. Top-$k$ full-pathway accuracy is the percentage of reactions with $R\leq k$, requiring every ground-truth elementary step along at least one reference pathway to rank within the top $k$.

\paragraph{Product prediction}
For product prediction, we sample complete pathways from reactants through intermediates to final products. We canonicalize the final products, group equivalent product predictions, and rank them by frequency. Top-$k$ product accuracy is the percentage of reactions whose ground-truth final product appears among the top $k$ ranked products. We use independent linear trajectories for \textsc{Bookkeeper} generation and compare it against beam search strategies used by FlowER and MechSMILES.

These three prediction tasks test complementary aspects of mechanistic modeling. Elementary-step accuracy measures recovery of the next mechanistic state; full-pathway accuracy assesses recovery of all steps along at least one reference pathway per reaction under a rank threshold; and product accuracy measures the recovery of the final product from sampled pathways. On the reaction-size-balanced subsets (Table~\ref{tab:balanced_performance}), \textsc{Bookkeeper} achieves the highest elementary-step top-1 accuracy (86.9\%), while FlowER performs better at top-5 and top-10. For pathway prediction, \textsc{Bookkeeper} slightly outperforms FlowER at top-1, whereas FlowER achieves higher accuracy at top-5 and top-10. In product prediction, \textsc{Bookkeeper} achieves the highest top-1 (59.9\%) and top-5 (89.1\%) accuracies, while FlowER leads at top-10. On the full test set (Table~\ref{tab:full_performance}), \textsc{Bookkeeper} achieves 90.72\% elementary-step top-1 accuracy versus 88.48\% for FlowER, but FlowER is higher at top-2 through top-10. For complete pathways, FlowER outperforms \textsc{Bookkeeper} at every reported rank. The contrasting rankings between the two test settings demonstrate that overall model performances are sensitive to the test set reaction-size distribution. We further attribute the low reaction pathway accuracy of \textsc{Bookkeeper} (Table~\ref{tab:full_performance}) to the sequential generation of complete pathways rather than through a beam search or top-p sampling.

\begin{table}[t]
\centering
\caption{Top-$k$ elementary-step, full-pathway, and product prediction accuracies (\%) on the balanced test set. Path accuracy uses the best route per reaction under the worst-step rank criterion, following FlowER. Product accuracy is the percentage of reactions whose ground-truth final product appears among the top \(k\) ranked products.}
\label{tab:balanced_performance}
\small
\begin{tabular}{l ccc ccc ccc}
\toprule
& \multicolumn{3}{c}{Step accuracy (\%)} & \multicolumn{3}{c}{Path accuracy (\%)} & \multicolumn{3}{c}{Product accuracy (\%)} \\
Model & Top-1 & Top-5 & Top-10 & Top-1 & Top-5 & Top-10 & Top-1 & Top-5 & Top-10 \\
\midrule
MechSMILES T5 & 68.3 & 82.6 & 82.8 & - & - & - & 47.6 & 77.4 & 78.2 \\
FlowER & 80.3 & \bf 98.8 & \bf 99.1 & 81.9 & \bf 95.7 & \bf 95.7 & 42.7 & 87.5 & \textbf{95.3} \\

\textsc{Bookkeeper} & \textbf{86.9} & 93.4 & 93.7 & \bf 82.9 & 85.1 & 85.1 & \textbf{59.9} & \textbf{89.1} & 90.1 \\
\bottomrule
\end{tabular}
\end{table}

\begin{table}[t]
\centering
\caption{Top-$k$ elementary-step and full-pathway prediction accuracies (\%) on the full FlowER test set. Path accuracy uses the best route per reaction under the worst-step rank criterion, following FlowER. FlowER results are quoted from its paper \cite{joung_electron_2025}.}
\label{tab:full_performance}
\small
\begin{tabular}{l cccc cccc}
\toprule
& \multicolumn{4}{c}{Step accuracy (\%)} & \multicolumn{4}{c}{Path accuracy (\%)} \\
Model & Top-1 & Top-2 & Top-5 & Top-10 & Top-1 & Top-2 & Top-5 & Top-10 \\
\midrule
 FlowER & 88.48 & {96.42} & {98.60} & {98.82} & {88.97} & {94.75} & {97.39} & {97.70} \\
 \textsc{Bookkeeper} & {90.72} & 94.08 & 95.11 & 95.37 & 88.48 & 93.37 & 96.14 & 96.26 \\
\bottomrule
\end{tabular}
\end{table}

\paragraph{Inference speed} We also compare elementary-step inference speed as a function of reaction size. We measure speed only at the elementary-step level so that differences in pathway-level search algorithms do not confound the comparison. All models are evaluated on the same dataset using the same NVIDIA RTX 4090 GPU and the same software environment. We group test examples by the number of atoms in the input reaction state and report average sampling time per elementary-step prediction. As shown in Figure \ref{fig_2} (a), \textsc{Bookkeeper} is consistently faster than FlowER and MechSMILES T5 across reaction sizes, reflecting the efficiency of directly sampling discrete electron movements rather than generating continuous BE-matrix trajectories or textual mechanism strings.  

\subsection{Out-of-distribution generalization and data-efficient fine-tuning}
We evaluate whether \textsc{Bookkeeper} generalizes to reaction types excluded from training and whether it can adapt to these new regimes with limited data (Appendix \ref{app:ood}). We first evaluate \textsc{Bookkeeper} trained on only in-distribution data directly on the OOD test reactions, without any reaction-type-specific training. We compare against FlowER without fine-tuning on the same OOD splits. This evaluates zero-shot or direct-transfer generalization to unseen reaction classes. After fine-tuning \textsc{Bookkeeper} as described in Section \ref{Methods}, each fine-tuned model is evaluated again on the same held-out OOD test set.

\begin{figure}[t]
  \centering
  \makebox[\textwidth][c]{%
    \includegraphics[width=1\linewidth]{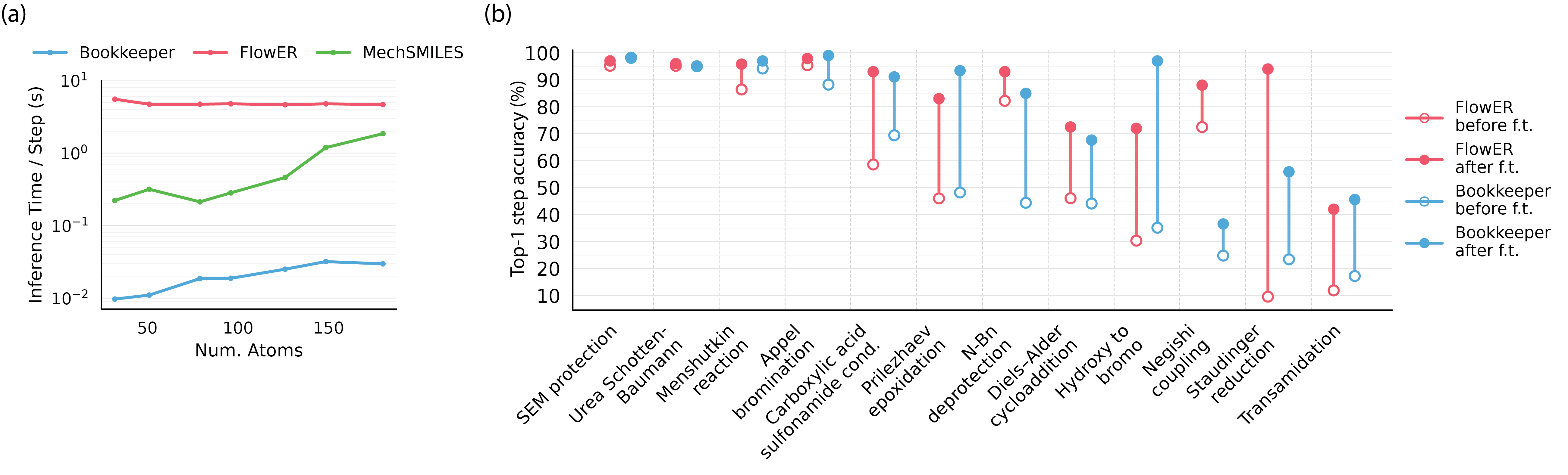}%
}{\vspace{-0.8em}}
  \caption{
  Inference speed and out-of-distribution fine-tuning of \textsc{Bookkeeper} compared with benchmark models. 
(a) Mean elementary-step inference time as a function of the number of atoms in the reacting molecule set.
(b) Top-1 elementary-step accuracy on 12 out-of-distribution reaction types before and after fine-tuning.
}
  \label{fig_2}
\vspace{-1em}
\end{figure} 

Figure~\ref{fig_2}(b) shows that \textsc{Bookkeeper} generalizes competitively to held-out reaction types before fine-tuning, achieving 56.9\% average top-1 accuracy and outperforming FlowER on 7 of 12 classes. With only 32 fine-tuning reactions per type, its average top-1 accuracy increases to 80.1\%, improving on 10 of 12 classes. These results suggest that \textsc{Bookkeeper} learns reusable electron-movement motifs that can be adapted to new reaction regimes with limited supervision, though certain reaction classes (e.g., Negishi coupling) may require broader coverage of organometallic or reagent-specific chemistry.

\subsection{Combinatorial screening for potential reaction discovery}
As an exploratory application, we use \textsc{Bookkeeper} for virtual screening of reagent-substrate combinations and generation of mechanistic hypotheses not represented by templates in the training data. Taking inspiration from the combinatorial reaction prediction setting used by Chen et al. \cite{chen_predicting_2025}, we identify 315 distinct reagent sets from 1,200 expert-curated reaction templates from the mechanistic dataset used in this work \cite{joung_reproducing_2024}, and then isolate 387 substrate sets from the test set by removing reagents from each reaction. Pairing every substrate set with every reagent set yields 121,905 combinations. For each combination, \textsc{Bookkeeper} generates reaction pathways using frequency-ranked stochastic sampling. The predicted reactions are then categorized into three classes, following the spirit of prior combinatorial reaction discovery analyses \cite{chen_predicting_2025}: (i) no reactivity, where the model fails to generate a chemically valid transformation; (ii) known reactivity, where the predicted mechanism matches a transformation already in the FlowER templates; and (iii) new or unseen reactivity, where the predicted mechanism corresponds to an unseen reagent-substrate transformation not in the FlowER templates.

\begin{figure}[t]
  \centering
  \makebox[\textwidth][c]{%
    \includegraphics[width=0.8\linewidth]{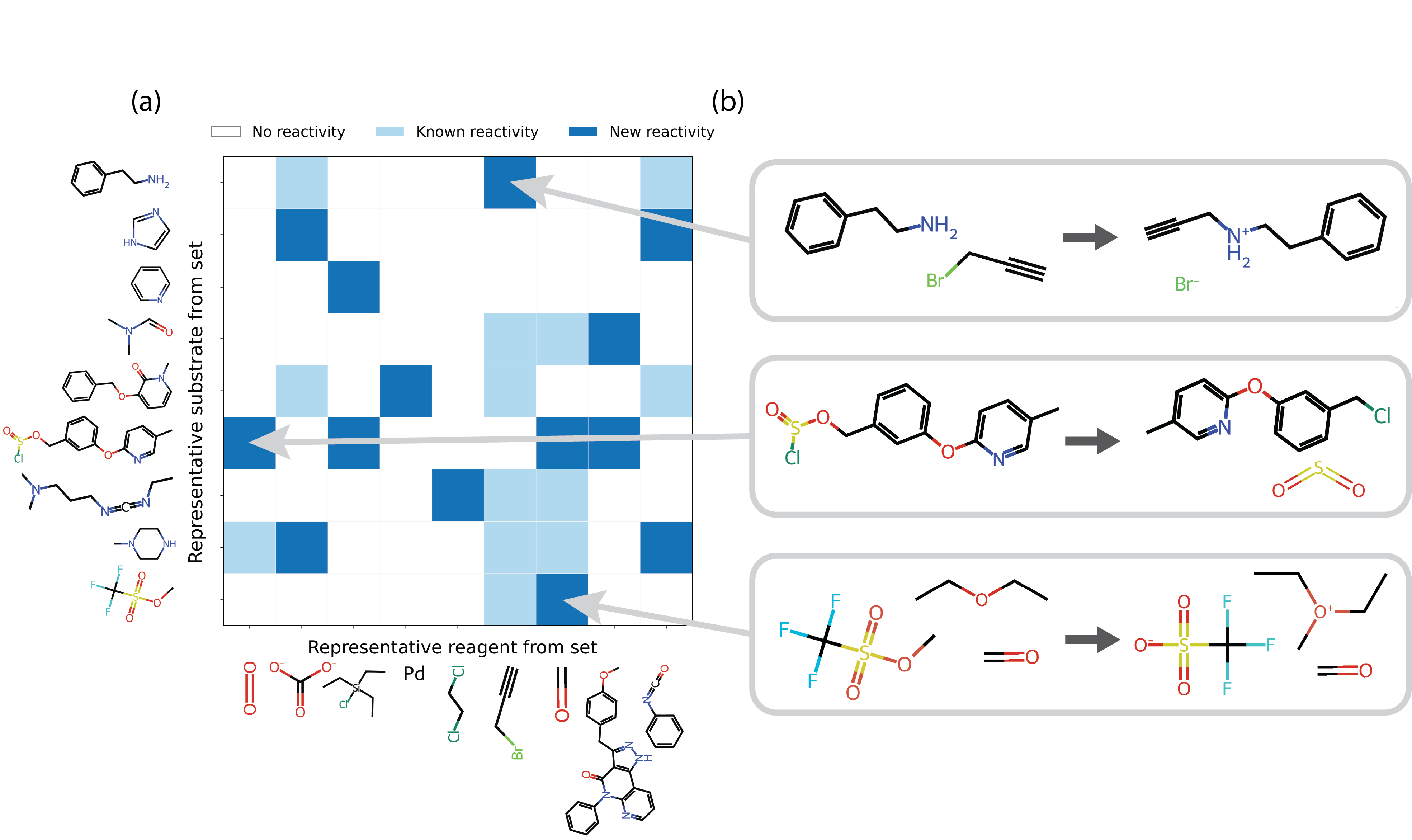}%
}
  \caption{ Combinatorial putative reaction discovery using \textsc{Bookkeeper}. 
(a) Representative subset of the reaction matrix, with each cell corresponding to one \emph{set} of reagents or substrates. 
(b) Details of representative predicted new overall transformations identified from the reaction matrix. ``New reactivity'' is defined as a reaction that does not correspond to any of the FlowER templates, rather than chemistry that is truly new.}
  \label{fig_3}
\vspace{-1.5em}
\end{figure}

Across 121,905 reagent-substrate combinations, \textsc{Bookkeeper} predicts 10,575 possible reactive combinations, corresponding to 8.7\% of the search space. Of these, 479 are new transformations, representing 4.5\% of all predicted reactive combinations and 0.4\% of all combinations. This sparsity is expected, as most arbitrary reagent-substrate pairings should not react, and the model should not force products for implausible inputs. Figure \ref{fig_3} shows that \textsc{Bookkeeper} can both recover known mechanistic analogs and identify a set of candidate new transformations. Because each prediction includes an explicit electron-movement trajectory, these candidates can be inspected mechanistically and prioritized for expert review or experimental validation.

\subsection{Learned chemical principles}
\paragraph{Regioselectivity} We next examine whether \textsc{Bookkeeper} learns chemically meaningful behavior beyond exact-match prediction by investigating its regioselectivity. Regioselective reactions are a direct probe of chemical context because although multiple sites may be formally reactive, steric, electronic, or directing effects often favor one site. Prior work has designed specialized machine learning models to account for nuanced regioselectivity for complex substrates \cite{ree_regioml_2022, schleinitz_designing_2025, pesciullesi_transfer_2020}. 

\begin{figure}[t]
  \centering
  \makebox[\textwidth][c]{%
    \includegraphics[width=1\linewidth]{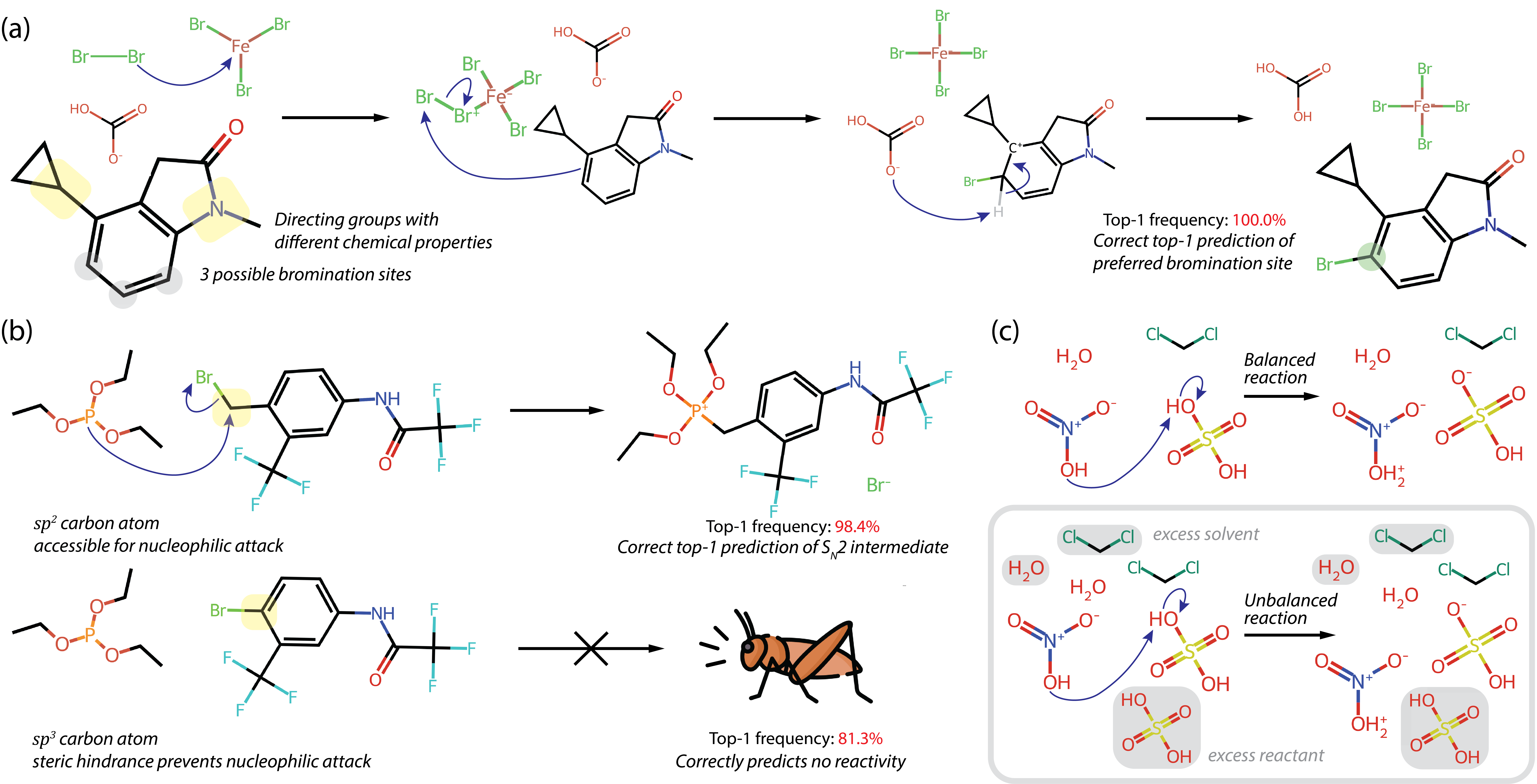}%
}
  \caption{
  \textsc{Bookkeeper} learns chemically meaningful behavior from reaction mechanisms. 
(a) Regioselective prediction in a substrate with three possible bromination sites for electrophilic aromatic substitution.
(b) Site-selective reactivity prediction for a Michaelis-Arbuzov reaction with different carbon hybridizations.
(c) Example of balanced vs. unbalanced reactions with imperfect stoichiometry and excess molecules.
  }
  \label{fig_4}
\end{figure}

We examine \textsc{Bookkeeper}'s predictions on representative examples, and Figure \ref{fig_4} shows that \textsc{Bookkeeper} indeed assigns high probability to chemically preferred reactive sites. In (a), the substrate contains three plausible bromination sites, but the model's top-ranked prediction exclusively selects the experimentally preferred site. This behavior could be indicative of the model utilizing the surrounding substituent context, including directing effects and electronic differences across the aromatic ring. Figure \ref{fig_4} (b) probes site selectivity in a Michaelis-Arbuzov reaction with competing carbon centers. \textsc{Bookkeeper} predicts nucleophilic attack at the accessible sp\(^2\) carbon with high frequency, generating the correct intermediate, while the sterically hindered sp\(^3\) carbon is assigned a no-reactivity outcome. This demonstrates the model's ability to suppress reactivity when the broader molecular environment makes a nominally possible attack unfavorable. These examples suggest that \textsc{Bookkeeper}'s electron-movement policy captures context-specific chemical preferences, which are essential for reaction mechanism prediction.

\paragraph{Imperfect stoichiometry and excess reagents}
A practical limitation of mass-conserved mechanistic models is that all reacting species must be present in the reactant pool, often at the exact molecular ratio required by the reaction. Indeed, prior formulations such as FlowER often assume that the input contains the exact number of molecules participating in the reaction \cite{joung_electron_2025}. In screening settings, however, the exact molecular ratios of reagents, solvents, additives, or even substrates may be unknown. It is thus useful for a mechanistic model to be robust against excess reactants, preserving nonparticipating species while still predicting the same mechanistic transformation among the reactive components. We test this by increasing the number of reagent and solvent molecules for elementary-step inputs and measuring whether top-\(k\) exact-match accuracy deteriorates. Specifically, we test a fixed perturbation level with 4 additional copies of a selected nonparticipating reagent or solvent appended, each with distinct atom-map indices. The original reacting species remain unchanged. The results show that \textsc{Bookkeeper} remains robust under this setting, with an elementary-step top-1 accuracy of 84.3\%, top-5 of 92.2\%, and top-10 of 92.7\%. This corresponds to a 2.6-point reduction at top-1 and a 1.0-point reduction at top-10 relative to the original inputs. Thus, even when the input contains additional molecules, the model retains nearly the same frequency-ranked prediction performance as in the perfectly specified setting.

Taken together, these experiments provide complementary evidence that \textsc{Bookkeeper} has learned some chemically meaningful behavior beyond benchmark exact-match accuracy. The regioselectivity cases show that the model can distinguish favored and disfavored reaction sites using molecular context, while the excess-species experiments show that it can avoid spurious reactivity when additional molecules are present. These properties support the use of \textsc{Bookkeeper} for reaction discovery and screening, where models must evaluate imperfectly specified reactant pools, propose mechanistically interpretable transformations, and avoid overreacting to chemically irrelevant species.

\section{Limitations}
\textsc{Bookkeeper} predicts symbolic arrow-pushing hypotheses rather than experimentally validated mechanisms or energy landscapes, and as such, conservation of atoms and electrons does not necessarily imply kinetic or thermodynamic feasibility. The model does not explicitly capture excited states, stereochemistry, or any three-dimensional considerations explicitly. Pathway accuracy depends on the sample budget, decoding temperature, and canonicalization. For screening new and unseen reactivity, candidate reactions require further independent literature, computational, and experimental validation. The regioselectivity results are illustrative examples rather than an aggregate selectivity benchmark.

The model's generalization is also ultimately bounded by the coverage and fidelity of available mechanistic training data. Rare mechanisms, poorly represented reagent classes, and chemistry outside templated reactions may require broader datasets or more targeted fine-tuning. The full-test-set comparison reveals  lower overall pathway accuracy than FlowER despite the more favorable top-1 ranking on reaction-size-balanced subsets. Therefore, our model's performance against benchmarks also depend on the evaluation distribution. 

\section{Conclusions}
We introduced \textsc{Bookkeeper}, which uses a discrete electron-bookkeeping formulation to preserve atoms and electrons by construction while generating interpretable arrow-pushing trajectories. It enables efficient mechanism generation without requiring continuous flow integration or beam search. Across the evaluated reaction sizes, the model achieves substantially faster elementary-step inference than FlowER and MechSMILES T5, highlighting its computational advantages for large-scale mechanistic reaction modeling. \textsc{Bookkeeper} improves top-ranked elementary-step prediction and pathway recovery on reaction-size-balanced benchmarks, while mostly matching FlowER on full-test-set results. \textsc{Bookkeeper} also adapts to held-out reaction types with limited data and produces mechanistic hypotheses that can be inspected for reaction screening. Importantly, its efficient sampling makes it particularly promising for applications requiring evaluation of large numbers of reagent-substrate combinations, where inference cost can be a major practical bottleneck. As mechanistic reaction datasets expand, the combination of explicit electron conservation, efficient autoregressive generation, and scalable candidate screening provides a promising foundation for computational exploration of chemical reaction space.

\section*{Acknowledgments}
This work was supported by the Machine Learning for Pharmaceutical Discovery and Synthesis Consortium and the National Science Foundation under grant no. CHE-2144153. S.L. received additional funding from the MIT Health and Life Sciences Collaborative (HEALS).

{
\small
\bibliographystyle{plainnat}
\bibliography{references}
}

\clearpage
\appendix

\section{Technical appendices and supplementary material}
\subsection{Code and data availability}
We include the code and instructions to access the mechanistic dataset \cite{joung_electron_2025} at \url{https://github.com/scaoyx/bookkeeper}.

\subsection{Mechanistic data}
\label{app:data}
We train \textsc{Bookkeeper} on elementary steps derived from applying templates to experimentally reported reactions, as curated by \citet{joung_reproducing_2024}, using the same splits as FlowER \cite{joung_electron_2025}. Each training example corresponds to one state transition \(S_t \rightarrow S_{t+1}\), where \(S_t=(A,E_t)\) and \(S_{t+1}=(A,E_{t+1})\) share the same atom set but differ in their electron configuration. The goal of the model is to generate a set of arrows that transforms \(E_t\) into \(E_{t+1}\).

Each elementary step \(\tau_i\) is converted from kekulized SMILES strings into a structure and electron bags. We compute the multiset difference between electron bags \(\Delta^- = E_t \setminus E_{t+1}\) and \(\Delta^+ = E_{t+1} \setminus E_t\), and construct ground truth arrows by matching each removed electron location in \(\Delta^-\) to an added electron location in \(\Delta^+\), yielding arrows \(\{u_i\} = \{(\ell^{(i)}_{\mathrm{src}},\ell^{(i)}_{\mathrm{dst}})\}\). The matching is done by constructing source-destination pairs that share the atom through which the electron is transferred, e.g., a lone electron on atom \(i\) moving to bond \((i,j)\), or a bonding electron \((i,j)\) moving to a lone location on one of its endpoint atoms, with deterministic tie-breaking when there are multiple identical electron locations. Trivial resonance moves within the same aromatic ring are removed.

\begin{figure}[H]
    \centering
    \includegraphics[width=0.7\linewidth]
    {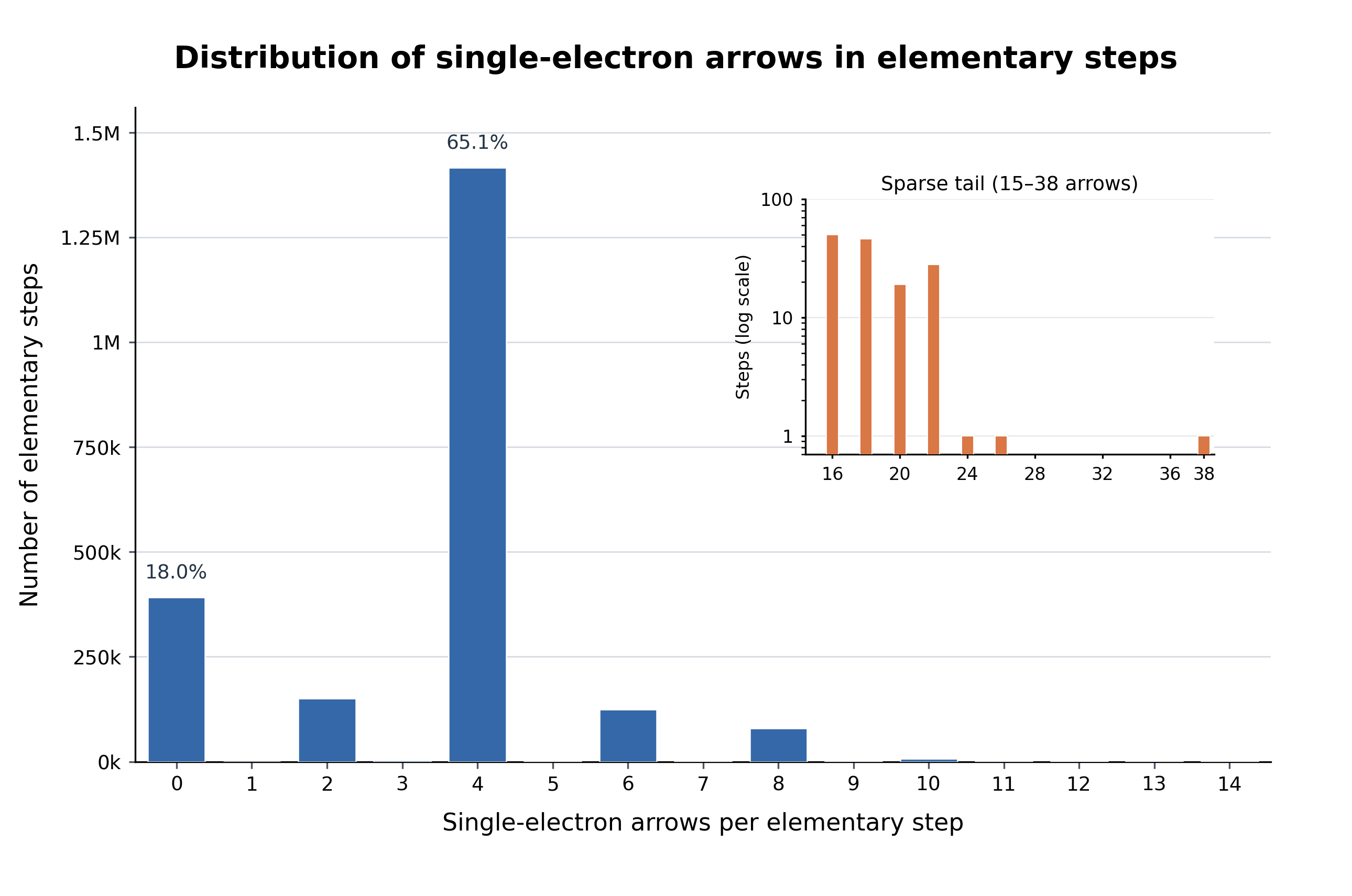}
    \caption{Distribution of the number of single-electron arrows per elementary step in the mechanistic corpus. The inset shows the sparse long tail spanning 15-38 arrows on a logarithmic count scale.}
    \label{fig:arrow_distribution}
\end{figure}

We further characterize the number of single-electron arrows associated with each elementary step in the mechanistic dataset. Figure~\ref{fig:arrow_distribution} shows the resulting distribution. The modal elementary step contains four single-electron arrows, accounting for 65.1\% of all steps. The 18.0\% containing no electron movements are the products-to-products static steps used to represent stopping conditions for the model to learn.

\subsection{Architecture, optimization, and training}
\label{app:training}
The \textsc{Bookkeeper} encoder has 8 bidirectional transformer layers with 8 attention heads, 512 hidden dimensions, 2048 feed-forward dimensions, and dropout 0.1, totaling 27.42 million trainable parameters. The model supports up to 384 atom tokens. Training uses AdamW with a learning rate of $3\times10^{-4}$, 100 optimizer-step linear warm-up followed by a constant learning rate, global-norm gradient clipping at 100, and 4 GPUs with 64 examples per GPU (global batch size 256). The training seed is 2026, validation occurs every 1,000 optimizer steps, and training lasted 167 epochs (422,458 optimizer steps, approximately 4 days 19 hours).

\subsection{Out-of-distribution fine-tuning}
\label{app:ood}
We evaluate data-efficient adaptation on the same 12 out-of-distribution (OOD) reaction types and data splits used by FlowER \cite{joung_electron_2025}. Starting from a \textsc{Bookkeeper} checkpoint pre-trained only on the in-distribution training set, we fine-tune a separate model for each held-out reaction type. For each type, fine-tuning uses only 32 overall reactions from the corresponding OOD training split, while evaluation is performed on the held-out OOD test set. This protocol measures whether a broadly trained electron-movement policy can be adapted to a new mechanistic regime with minimal reaction-type-specific supervision, rather than requiring full retraining on a large specialized dataset.

\subsection{Evaluation protocol}
\label{app:uncertainty}
For evaluation, we generate 64 raw candidate predictions per test input for each compared method while retaining each model's native generators. Successful outputs are reconstructed and canonicalized. Equivalent molecular outcomes or complete intermediate sequences are grouped by chemical identity, and unique predictions are ranked by descending sample frequency. Invalid, missing, or timed-out raw outputs remain failures and are not replaced. Tables~\ref{tab:step_accuracy} and~\ref{tab:pathway_accuracy} report the reaction-size-balanced subset results, with uncertainty estimated by standard deviations from repeated sampling of such balanced subsets. The full test set comparison (Table~\ref{tab:full_performance}) uses all available test examples and is reported as point estimates without bootstrap error bars. Note that the following results are obtained by running the original FlowER model, not FlowER-large, which was the one evaluated in Section \ref{Experiments}.

\begin{table}[H]
\centering
\caption{Top-$k$ elementary-step prediction accuracy (\%) on reaction-size-balanced test subsets with standard deviations.}
\label{tab:step_accuracy}
\small
\begin{tabular}{lcccc}
\toprule
Model & Top-1 & Top-2 & Top-5 & Top-10 \\
\midrule
MechSMILES T5
& $68.3\pm1.47$ & $77.7\pm1.08$
& $82.6\pm0.79$ & $82.8\pm0.77$ \\
FlowER
& $69.2\pm1.58$ & $80.2\pm1.51$
& $84.6\pm1.41$ & $84.8\pm1.41$ \\
\textsc{Bookkeeper}
& $\mathbf{86.9\pm1.15}$ & $\mathbf{90.7\pm0.89}$
& $\mathbf{93.4\pm0.81}$ & $\mathbf{93.7\pm0.80}$ \\
\bottomrule
\end{tabular}
\end{table}

\begin{table}[H]
\centering
\caption{Top-$k$ reaction pathway prediction accuracy (\%) on reaction-size-balanced test subsets with standard deviations.}
\label{tab:pathway_accuracy}
\small
\begin{tabular}{lcccc}
\toprule
Model & Top-1 & Top-2 & Top-5 & Top-10 \\
\midrule
MechSMILES T5
& $47.6\pm2.58$ & $62.2\pm1.68$
& $77.4\pm0.84$ & $78.2\pm0.84$ \\
FlowER
& $42.7\pm4.01$ & $76.0\pm3.25$
& $87.5\pm1.63$ & $\mathbf{95.3\pm1.42}$ \\
\textsc{Bookkeeper}
& $\mathbf{59.9\pm3.93}$ & $\mathbf{79.7\pm2.72}$
& $\mathbf{89.1\pm2.13}$ & $90.1\pm2.06$ \\
\bottomrule
\end{tabular}
\end{table}

\subsection{Shuffled-prefix supervision ablation}
\label{app:shuffle_ablation}
To evaluate the contribution of shuffled-prefix supervision, we compare \textsc{Bookkeeper} against a fixed-order training variant without shuffled-prefix augmentation. All other experimental settings are held constant. Shuffled-prefix supervision exposes the model to different valid orderings of electron movements within an elementary step, encouraging it to predict the remaining movements from arbitrary partially updated states rather than relying on a predetermined arrow order.

Table~\ref{tab:shuffle_ablation} reports the top-$k$ prediction accuracies for both variants. Shuffled-prefix supervision consistently improves performance across all evaluated $k$. For reaction pathway
prediction, top-1 accuracy increases from 54.2\% to 59.9\%, while top-10 accuracy increases from 87.9\% to 90.1\%. The effect is considerably larger in the elementary step evaluation, where top-1 accuracy increases from 16.1\% to 88.6\% and top-10 accuracy increases from 64.6\% to 94.3\%.

These results demonstrate the importance of accounting for the permutation invariance of electron movements within an elementary step. Training on a single fixed arrow ordering can introduce dependencies on an arbitrary serialization, whereas shuffled-prefix supervision allows the model to learn from a broader range of chemically equivalent partial-step configurations.

\begin{table}[H]
\centering
\caption{Ablation of shuffled-prefix supervision. Top-$k$ prediction accuracy (\%) with and without shuffled-prefix training. All other experimental settings are held constant.}
\label{tab:shuffle_ablation}
\small
\setlength{\tabcolsep}{5pt}
\begin{tabular}{llcccc}
\toprule
Task & Training & Top-1 & Top-2 & Top-5 & Top-10 \\
\midrule
\multirow{2}{*}{Reaction pathway}
& Shuffled-prefix
& \textbf{59.9} & \textbf{79.7} & \textbf{89.1} & \textbf{90.1} \\
& No shuffle
& 54.2 & 77.1 & 86.2 & 87.9 \\
\midrule
\multirow{2}{*}{Elementary step}
& Shuffled-prefix
& \textbf{88.6} & \textbf{92.6} & \textbf{94.2} & \textbf{94.3} \\
& No shuffle
& 16.1 & 55.2 & 63.5 & 64.6 \\
\bottomrule
\end{tabular}
\end{table}

\subsection{Inference efficiency and robustness experiments}
\label{app:efficiency_robustness}
\paragraph{Inference efficiency across reaction sizes}
We examine elementary-step inference time as a function of the number of atoms in the input reaction state.
Table~\ref{tab:inference_timing} presents timing measurements for \textsc{Bookkeeper}, FlowER, and MechSMILES T5 across ten reaction-size groups. The measurements are obtained at the elementary-step level rather than for end-to-end reaction pathway generation, avoiding differences in pathway-level search procedures.

\textsc{Bookkeeper} achieves lower elementary-step inference times than both baselines across all evaluated reaction sizes. Its measured inference time increases from 9.67 ms for reactions with an average of 32 atoms to 47.88 ms for reactions with an average of 261 atoms. In contrast, FlowER requires approximately 93-111 ms across these groups, while MechSMILES T5 exhibits substantially higher and more variable inference times, particularly for larger reaction systems.

\begin{table}[H]
\centering
\caption{Elementary-step inference time (ms) across reaction-size groups. Reaction size is measured by the mean number of atoms in each group.}
\label{tab:inference_timing}
\small
\setlength{\tabcolsep}{7pt}
\begin{tabular}{rccc}
\toprule
Mean atoms
& \textsc{Bookkeeper}
& FlowER
& MechSMILES T5 \\
\midrule
32  & $9.67\pm0.02$  & $110.70\pm60.73$ & $221.37\pm1.01$ \\
51  & $10.98\pm0.08$ & $94.23\pm0.96$   & $316.37\pm2.21$ \\
79  & $18.56\pm0.31$ & $94.43\pm0.19$   & $212.33\pm2.20$ \\
96  & $18.71\pm0.21$ & $95.55\pm1.75$   & $282.26\pm5.07$ \\
126 & $24.99\pm0.23$ & $92.66\pm0.50$   & $460.17\pm99.03$ \\
148 & $31.82\pm0.70$ & $95.50\pm3.06$   & $1188.75\pm353.51$ \\
180 & $29.61\pm0.05$ & $93.21\pm0.46$   & $1851.50\pm13.14$ \\
204 & $40.85\pm0.91$ & $95.65\pm1.71$   & $586.73\pm5.71$ \\
223 & $42.78\pm0.33$ & $94.91\pm0.26$   & $1431.76\pm63.59$ \\
261 & $47.88\pm0.30$ & $95.49\pm0.52$   & $1926.87\pm126.80$ \\
\bottomrule
\end{tabular}
\end{table}

\paragraph{Robustness to excess reagents and solvents}
We evaluate the robustness of \textsc{Bookkeeper} to imperfectly specified reactant stoichiometry by introducing additional copies of nonparticipating reagents or solvents into elementary-step inputs. For each perturbed reaction, 4 additional copies of the selected nonparticipating species are appended, giving five copies in total. The reacting components remain unchanged, and the appended copies receive unique atom-mapping indices. Exact-match evaluation requires the additional molecules to remain unchanged in the predicted state.

Table~\ref{tab:excess_reagents} compares elementary-step prediction accuracy on the original and perturbed inputs. Adding excess species reduces top-1 accuracy from 86.9\% to 84.3\% and top-10 accuracy
from 93.7\% to 92.7\%, corresponding to decreases of 2.6 and 1.0 percent, respectively. These results indicate that \textsc{Bookkeeper} can preserve nonparticipating molecules while recovering the intended transformation with relatively small accuracy degradation at the tested perturbation level.

\begin{table}[H]
\centering
\caption{Elementary-step top-$k$ accuracy (\%) with the original reactant pool and with 4 additional copies of nonparticipating reagents or solvents.}
\label{tab:excess_reagents}
\small
\setlength{\tabcolsep}{5pt}
\begin{tabular}{lcccccccccc}
\toprule
Setting & Top-1 & Top-2 & Top-3 & Top-4 & Top-5
& Top-6 & Top-7 & Top-8 & Top-9 & Top-10 \\
\midrule
Original
& 86.9 & 90.7 & 92.8 & 93.2 & 93.4
& 93.6 & 93.6 & 93.6 & 93.7 & 93.7 \\
Excess
& 84.3 & 88.1 & 90.3 & 91.2 & 92.2
& 92.3 & 92.3 & 92.4 & 92.7 & 92.7 \\
\bottomrule
\end{tabular}
\end{table}

\subsection{Endpoint ordering and predicted stop actions}
\label{app:order_stop}
\textsc{Bookkeeper}'s shuffled-prefix training featurizer draws a random orientation for two-atom electron locations, including the destination of each remaining movement. In the molecular representation, an electron shared by atoms $i$ and $j$ is equal to one shared by $j$ and $i$, independent of orientation. 

During elementary-step sampling, chemical integrity checks test atoms appearing in sampled arrows and confirm that each must have a nonzero associated electron count, an even number of bonding electrons, and total incident occupancy no greater than the configured valence-shell capacity. The candidate is accepted immediately when these conditions hold. Furthermore, trajectories containing both a move and its reverse are rejected, whether or not the moves are consecutive. In the linear full-trajectory sampler, no-move, reverse-loop detection, or the maximum number of iterations ends a trajectory. The integrity check is also used as a method to identify elementary steps in the pathway.

\subsection{Full pathway evaluation}
\label{app:pathway_eval}
We evaluate full reaction pathway prediction by requiring exact recovery of each ground-truth elementary step in a pathway, rather than only the final reaction products. Starting from the reactants, \textsc{Bookkeeper} autoregressively generates individual electron movements and updates the electron configuration after each action. Completed elementary steps are identified using the chemical integrity criteria described in Section~3.4 and Appendix~\ref{app:order_stop}, yielding a sequence of elementary steps from the initial reactants to the predicted products.

For each test reaction, we generate 64 independent stochastic trajectories using linear autoregressive sampling rather than beam search. The same raw-output budget and post-generation ranking procedure are applied to all benchmark models, while each retains its native generation algorithm. Generated outputs are canonicalized, grouped by equivalent predictions, and ranked according to their empirical frequencies among the 64 samples. Invalid, incomplete, timed-out, or missing predictions are not replaced by additional samples, and all test examples remain in the evaluation denominator.

\subsection{Sampling temperature sensitivity}
\label{app:temperature}
We investigate the sensitivity of \textsc{Bookkeeper} to sampling temperature by varying the temperature while maintaining a fixed budget of 64 raw predictions per test input. All other experimental settings remain unchanged across different temperatures. We evaluate temperatures of $0.7$, $1.0$, and $2.0$ for both
elementary-step prediction and complete reaction pathway generation.

Table~\ref{tab:temperature_step} presents elementary-step prediction accuracies. Performance remains relatively stable across the evaluated temperatures. Top-1 accuracy varies from 88.4\% to 89.0\%, while top-10 accuracy ranges from 93.9\% to 94.3\%. The largest difference across temperatures at any reported $k$ is 0.9 percentage points, suggesting that single-step prediction is relatively insensitive to temperature within the investigated range.

\begin{table}[H]
\centering
\caption{Elementary-step top-$k$ prediction accuracy (\%) at different sampling temperatures, using 64 samples per input.}
\label{tab:temperature_step}
\small
\setlength{\tabcolsep}{6pt}
\begin{tabular}{lccc}
\toprule
Top-$k$ & $T=0.7$ & $T=1.0$ & $T=2.0$ \\
\midrule
1  & \textbf{89.0} & 88.6 & 88.4 \\
2  & \textbf{92.7} & 92.6 & 92.4 \\
3  & 93.2 & \textbf{93.6} & 92.7 \\
4  & 93.4 & \textbf{93.9} & 93.2 \\
5  & 93.4 & \textbf{94.2} & 93.6 \\
6  & 93.6 & \textbf{94.3} & 93.8 \\
7  & 93.6 & \textbf{94.3} & 94.0 \\
8  & 93.7 & \textbf{94.3} & 94.0 \\
9  & 93.9 & \textbf{94.3} & 94.0 \\
10 & 93.9 & \textbf{94.3} & 94.0 \\
\bottomrule
\end{tabular}
\end{table}

In contrast, complete pathway prediction exhibits greater sensitivity to sampling temperature, as shown in
Table~\ref{tab:temperature_pathway}. At $T=0.7$, \textsc{Bookkeeper} achieves 59.9\% top-1 and 90.1\% top-10 accuracy. Increasing the temperature to $T=1.0$ reduces these values to 51.0\% and 87.5\%, respectively, while $T=2.0$ yields 57.3\% and 81.8\%. The lowest evaluated temperature achieves the highest pathway accuracy at every reported $k$.

\begin{table}[H]
\centering
\caption{Reaction pathway top-$k$ prediction accuracy (\%) at different sampling temperatures, using 64 samples per input.}
\label{tab:temperature_pathway}
\small
\setlength{\tabcolsep}{6pt}
\begin{tabular}{lccc}
\toprule
Top-$k$ & $T=0.7$ & $T=1.0$ & $T=2.0$ \\
\midrule
1  & \textbf{59.9} & 51.0 & 57.3 \\
2  & \textbf{79.7} & 78.6 & 76.6 \\
3  & \textbf{84.9} & 83.3 & 79.2 \\
4  & \textbf{87.0} & 84.4 & 80.2 \\
5  & \textbf{89.1} & 86.5 & 80.7 \\
6  & \textbf{89.6} & 87.0 & 81.3 \\
7  & \textbf{89.6} & 87.5 & 81.3 \\
8  & \textbf{89.6} & 87.5 & 81.8 \\
9  & \textbf{90.1} & 87.5 & 81.8 \\
10 & \textbf{90.1} & 87.5 & 81.8 \\
\bottomrule
\end{tabular}
\end{table}

The greater temperature sensitivity of pathway prediction is consistent with the accumulation of sampling differences over multiple autoregressive electron movements. Under a fixed sampling budget, lower temperatures concentrate probability mass on higher-probability trajectories, increasing their frequency in the resulting candidate ranking. Higher temperatures likely distribute samples more broadly across possible trajectories, which can reduce exact-match accuracy when predictions are ranked by empirical frequency.

Therefore, we demonstrate a tradeoff between concentrating predictions on high-probability outcomes and exploring alternative reaction trajectories. The preferred sampling temperature may thus depend on whether the application prioritizes exact pathway recovery or broader candidate generation.

\subsection{Forward major product prediction comparison}
\label{app:forward_prediction}
Although \textsc{Bookkeeper} is designed for mechanistic reaction prediction, the accuracy of its final reaction products provides an additional measure of practical utility. We therefore compare \textsc{Bookkeeper} with two established forward reaction prediction models, Graph2SMILES (G2S) \cite{tu2022permutation} and Chemformer \cite{irwin2022chemformer}, on the same major product prediction task.

As in the reaction-size-balanced mechanism evaluation in Section~\ref{Performance}, we partition the mechanistic test set into six bins by input atom count, spanning 50 to 384 atoms, and sample 168 reactions from each bin. Each evaluation sample therefore contains 1,010 reactions. This procedure reduces the dominance of smaller reactions in the original test distribution and provides greater representation of larger molecular systems.

Because conventional forward predictors generally target the major reaction product rather than the complete set of products and byproducts, we heuristically identify the major product for each reference reaction and evaluate all three models against this common target. Predictions are scored using exact major product matching, with invalid SMILES, failed predictions, and missing outputs retained in the evaluation denominator. This metric differs from the complete pathway criteria used elsewhere in this work.

Table~\ref{tab:forward_prediction} reports the resulting top-$k$ accuracies. \textsc{Bookkeeper} achieves 87.8\% top-1 and 94.7\% top-10 accuracy, compared with 79.7\% and 86.0\% for Chemformer and 64.4\% and 90.3\% for Graph2SMILES. \textsc{Bookkeeper} obtains the highest mean accuracy at every reported $k$. Chemformer outperforms Graph2SMILES at top-1, while Graph2SMILES achieves higher accuracy from top-4 onward.

\begin{table}[H]
\centering
\caption{Top-$k$ major product prediction accuracy (\%) on balanced test sets. Values are reported as mean $\pm$ standard deviation across evaluation samples.}
\label{tab:forward_prediction}
\small
\setlength{\tabcolsep}{6pt}
\begin{tabular}{lccc}
\toprule
Top-$k$ & \textsc{Bookkeeper} & Chemformer & Graph2SMILES \\
\midrule
1  & $\mathbf{87.8\pm3.9}$ & $79.7\pm3.2$ & $64.4\pm3.9$ \\
2  & $\mathbf{92.4\pm2.7}$ & $83.1\pm3.1$ & $77.8\pm2.7$ \\
3  & $\mathbf{93.4\pm2.4}$ & $84.5\pm2.4$ & $83.0\pm2.4$ \\
4  & $\mathbf{93.8\pm2.3}$ & $84.9\pm2.2$ & $87.4\pm2.3$ \\
5  & $\mathbf{93.8\pm2.3}$ & $85.3\pm2.0$ & $89.4\pm2.1$ \\
6  & $\mathbf{94.4\pm2.1}$ & $85.9\pm2.0$ & $90.3\pm2.1$ \\
7  & $\mathbf{94.6\pm2.1}$ & $86.0\pm2.0$ & $90.3\pm2.1$ \\
8  & $\mathbf{94.6\pm2.1}$ & $86.0\pm2.0$ & $90.3\pm2.1$ \\
9  & $\mathbf{94.7\pm2.1}$ & $86.0\pm2.0$ & $90.3\pm2.1$ \\
10 & $\mathbf{94.7\pm2.1}$ & $86.0\pm2.0$ & $90.3\pm2.1$ \\
\bottomrule
\end{tabular}
\end{table}

We note that the two forward-prediction baselines differ in their pretraining and fine-tuning histories. For Graph2SMILES, we use the baseline checkpoint previously employed by FlowER, initialized from a USPTO-full pretrained model and subsequently fine-tuned on the same mechanistic training distribution as \textsc{Bookkeeper}. For Chemformer, we use the publicly released checkpoint fine-tuned on USPTO-mixed, without additional fine-tuning on our mechanistic training set. Consequently, the comparison does not control for differences in training data, pretraining procedures, or model architecture. Potential overlap between the Chemformer training data and our evaluation subset also cannot be excluded.

The different checkpoint histories may partially explain the models' rank-dependent performance. Chemformer achieves substantially higher top-1 accuracy than Graph2SMILES, whereas Graph2SMILES recovers more reference products at larger $k$. Similar relative trends have been observed in previous comparisons of these architectures, although differences in benchmark distributions make direct attribution of the observed
gap to a particular modeling choice difficult.

Overall, these results indicate that the explicit mechanistic representation of \textsc{Bookkeeper} does not necessarily compromise major product prediction accuracy on the test set.

\subsection{Model development and negative results}
\label{app:negative_results}
The design of \textsc{Bookkeeper} emerged from exploring alternative formulations of mechanism-level reaction generation. In particular, we investigated whether the complete set of electron movements within an elementary step could be predicted in parallel, rather than generating individual movements autoregressively.

We developed a discrete denoising formulation that predicts the destination of every electron after each elementary step. Starting from the initial electron bag, a featurization module constructs per-electron destination targets together with move/stop indicators. During training, destination coordinates and movement indicators are corrupted, and the denoising model learns to reconstruct the complete post-step electron configuration. At inference time, all electron destinations are predicted through 100 denoising iterations.

We implemented transformer variants with 8 and 12 layers, both using a hidden dimension of 512. In these experiments, predictions were dominated by the no-move outcome; i.e., the model frequently predicted that nearly all electrons should retain their original locations. This behavior is consistent with the strong class imbalance between moving and stationary electrons in typical elementary reactions. Moreover, independently generated electron destinations could produce mutually inconsistent changes when interpreted as a simultaneous set of arrows.

These difficulties motivated the autoregressive formulation used in \textsc{Bookkeeper}. Rather than predicting the entire electron redistribution in parallel, the model selects one electron source, predicts its destination, applies the movement to the current state, and conditions subsequent predictions on the updated electron configuration. This sequential formulation allows each prediction to incorporate the consequences of previously generated movements.

Because the arrows within a single elementary step are chemically unordered, sequential generation also introduces a potential dependence on arbitrary arrow serialization. We address this through shuffled-prefix supervision, which exposes the model to multiple orderings and partial configurations of the same elementary step. The ablation in Appendix~\ref{app:shuffle_ablation} demonstrates the importance of this training strategy.

Another practical design choice is the use of copy indices for electron tokens. Multiple electrons can occupy the same location, such as a lone pair or a covalent bond. Assigning distinct computational indices allows the model to select, update, and track individual tokens while preserving the total electron count. These indices distinguish elements of the model representation and do not imply physically distinguishable electron identities.

Our final architecture uses a bidirectional transformer to encode the complete current atom structure and electron bag, and an autoregressive action factorization to predict successive electron movements. An alternative decoder-only transformer could also generate serialized electron-movement sequences, but would require an explicit treatment of the arbitrary ordering of simultaneous arrows. We did not train a matched decoder-only control. Accordingly, the present experiments motivate our architectural choices without necessarily establishing that \textsc{Bookkeeper}'s architecture is uniquely optimal.

\subsection{Template coverage and literature supporting new reactivity candidates}
\label{app:template_coverage}
\paragraph{Mechanistic template coverage} The mechanistic dataset is predominantly organic chemistry transformations, with comparatively limited representation of organometallic and transition-metal-catalyzed chemistry. To characterize this imbalance, we categorize the reaction templates according to their chemical transformation classes. Table~\ref{tab:template_coverage} summarizes the resulting distribution.

Of the 252 categorized templates, 207 (82.1\%) correspond to other organic reaction types, while the remaining 45 (17.9\%) span transition-metal-catalyzed cross-couplings, main-group organometallic transformations, organosilicon chemistry, and related reactions.

\begin{table}[H]
\centering
\caption{Distribution of chemical transformation classes in the mechanistic template catalog}
\label{tab:template_coverage}
\small
\setlength{\tabcolsep}{6pt}
\begin{tabular}{p{0.29\linewidth}p{0.39\linewidth}rr}
\toprule
Category & Representative reactions & Templates & Share (\%) \\
\midrule
Transition metal catalyzed cross-coupling
& Suzuki, Buchwald-Hartwig, Miyaura borylation
& 13 & 5.2 \\
Other cross-coupling
& Sonogashira, Stille, Kumada
& 11 & 4.4 \\
Main group organometallic
& Grignard
& 10 & 4.0 \\
Organosilicon
& Hosomi-Sakurai
& 6 & 2.4 \\
Other related organometallic transformations
& Barbier, Petasis, Weinreb ketone synthesis,
olefin metathesis
& 5 & 2.0 \\
Organic transformations
& Remaining organic reaction types
& 207 & 82.1 \\
\midrule
Total & & 252 & 100.0 \\
\bottomrule
\end{tabular}
\end{table}

This uneven coverage provides a plausible explanation for the comparatively weak performance observed when adapting \textsc{Bookkeeper} to Negishi cross-coupling reactions. Although the electron bookkeeping representation can describe electron redistribution involving metal-containing species, reliable prediction of such chemistry possibly depends on sufficient and representative mechanistic training data. The observed limitation therefore partially relates to the chemical coverage of the available corpus rather than an inherent inability of the representation to encode organometallic transformations.

\paragraph{Literature-supported candidate transformation} As an illustrative case from combinatorial reaction screening, \textsc{Bookkeeper} predicts the methylation of diethyl ether by methyl trifluoromethanesulfonate (methyl triflate, MeOTf), yielding a methyldiethyloxonium triflate species.

The proposed elementary step involves nucleophilic attack by the ether oxygen on the methyl group of MeOTf, accompanied by cleavage of the carbon-triflate bond. This transformation is consistent with an $S_\mathrm{N}2$-type methyl transfer and produces a trialkyloxonium cation with triflate as its counteranion. The associated carbon dioxide molecule in the screened reactant combination remains unchanged.

The predicted product belongs to the trialkyloxonium/Meerwein-type family of compounds, for which related species have been documented experimentally. In particular, trialkyloxonium chemistry reported by Staneke et al.\ (1996) \cite{staneke1996competing} provides literature support for the chemical plausibility of this prediction.

This example illustrates how \textsc{Bookkeeper} can propose mechanistically interpretable candidate transformations whose plausibility can be assessed against established chemical literature. However, we emphasize that a candidate transformation absent from the training templates is not necessarily a previously undiscovered reaction. Establishing chemical novelty and experimental feasibility requires additional literature analysis, expert evaluation, and computational or downstream experimental validation.

\end{document}